\documentclass{aa}

\usepackage{natbib}
\usepackage{multirow}
\usepackage{float}

\usepackage{graphicx}

\usepackage{txfonts}
\usepackage{comment}
\usepackage{placeins}
\usepackage{afterpage}
\usepackage{makecell}
\usepackage{subcaption}

\usepackage{hyperref}
\usepackage{multicol}
\usepackage[dvipsnames]{xcolor}

\begin{document}

   \title{Surveying the Whirlpool at Arcseconds with NOEMA (SWAN)}

   \subtitle{III. $^{13}$CO/C$^{18}$O ratio variations across the M51 galaxy}

      \author{I. Gali\'{c} \inst{\ref{AIfA}}
        \and Mallory Thorp \inst{\ref{AIfA}}
        \and Frank Bigiel\inst{\ref{AIfA}}
        \and Eva Schinnerer\inst{\ref{MPIA}}
        \and Jakob den Brok\inst{\ref{CfA}}
        \and Hao He \inst{\ref{AIfA}}
        \and Mar\'ia~J.~Jim\'enez-Donaire\inst{\ref{ESA},\ref{OAN}}
        \and Lukas Neumann \inst{\ref{ESO}}
        \and Jerome Pety\inst{\ref{IRAM},\ref{LERMA}}
        \and Sophia K. Stuber\inst{\ref{MPIA}}
        \and Antonio Usero\inst{\ref{OAN}}
        \and Ashley~T.~Barnes \inst{\ref{ESO}}
        \and Dario Colombo \inst{\ref{AIfA}}
        \and Daniel~A.~Dale \inst{\ref{UW}}
        \and Timothy A. Davis \inst{\ref{CHART}}
        \and J. E. M\'endez-Delgado \inst{\ref{UNAM}}
        \and Hsi-An Pan \inst{\ref{TKU}}
        \and Miguel ~Querejeta \inst{\ref{OAN}}
        \and Thomas G. Williams\inst{\ref{Ox}}
          }

      \institute{Argelander Institute for Astronomy (AIfA), University of Bonn,
              Auf dem Hügel 71, 53121 Bonn \label{AIfA}
        \and Max-Planck-Institut für Astronomie, Königstuhl 17, 69117 Heidelberg, Germany\label{MPIA}
        \and Center for Astrophysics $\mid$ Harvard \& Smithsonian, 60 Garden St., 02138 Cambridge, MA, USA\label{CfA}
        \and AURA for the European Space Agency (ESA), ESA Office, Space Telescope Science Institute, 3700 San Martin Drive, Baltimore, MD 21218, USA\label{ESA}
        \and Observatorio Astron\'omico Nacional (IGN), C/ Alfonso XII, 3, E-28014 Madrid, Spain\label{OAN}
        \and European Southern Observatory, Karl-Schwarzschild Stra{\ss}e 2, D-85748 Garching bei M\"{u}nchen, Germany \label{ESO}
        \and IRAM, 300 rue de la Piscine, 38400 Saint Martin d'H\`eres, France\label{IRAM}
        \and Sorbonne Universit\'e, Observatoire de Paris, Universit\'e PSL, \' Ecole normale sup\`erieure, CNRS, LERMA, F-75005, Paris, France\label{LERMA}
        \and Department of Physics and Astronomy, University of Wyoming, Laramie, WY 82071, USA\label{UW}
        \and Cardiff Hub for Astrophysics Research \& Technology, School of Physics \& Astronomy, Cardiff University, Queens Buildings, Cardiff CF24 3AA, UK\label{CHART}
        \and Instituto de Astronom\'ia, Universidad Nacional Aut\'onoma de M\'exico, Ap. 70-264, 04510 CDMX, Mexico \label{UNAM}
        \and Department of Physics, Tamkang University, No.151, Yingzhuan Road, Tamsui District, New Taipei City 251301, Taiwan \label{TKU}
        \and Sub-department of Astrophysics, Department of Physics, University of Oxford, Keble Road, Oxford OX1 3RH, UK\label{Ox}
        }

   \date{Accepted July 30, 2025; Received January 10, 2025}

  \abstract
   {CO isotopologues are common tracers of the bulk molecular gas in extragalactic studies, providing insights into the physical and chemical conditions of the cold molecular gas, a reservoir for star formation.}
   {Since star formation occurs within molecular clouds, mapping CO isotopologues at cloud-scale is important to understanding the processes driving star formation. However, achieving this mapping at such scales is challenging and time-intensive. The Surveying the Whirlpool Galaxy at Arcseconds with NOEMA (SWAN) survey addresses this by using the Institut de radioastronomie millimétrique (IRAM) NOrthern Extended Millimeter Array (NOEMA) to map the $^{13}$CO(1-0) and C$^{18}$O(1-0) isotopologues, alongside several dense gas tracers, in the nearby star-forming galaxy M51 at high sensitivity and spatial resolution ($\approx$ 125 pc).}
   {We examine the $^{13}$CO(1-0) to C$^{18}$O(1-0) line emission ratio as a function of galactocentric radius and star formation rate surface density to infer how different chemical and physical processes affect this ratio at cloud scales across different galactic environments: nuclear bar, molecular ring, northern and southern spiral arms.}
   {In line with previous studies conducted at kiloparsec scales for nearby star-forming galaxies, we find a moderate positive correlation with galactocentric radius and a moderate negative correlation with star formation rate surface density across the field-of-view (FoV), with slight variations depending on the galactic environment.}
   {We propose that selective nucleosynthesis and changes in the opacity of the gas are the primary drivers of the observed variations in the ratio.}


   \maketitle

\section{Introduction}
Unravelling the physical and chemical characteristics of the interstellar medium (ISM) has been instrumental in understanding the mechanisms driving the evolution of galaxies to their current states. In simple terms, under cold \citep[10~K - 100~K;][]{1997Wilson,2017Tang} and relatively dense \citep[$\approx$ $10^2$ cm$^{-3}$;][]{2015shirley} conditions the ISM material clumps up to form giant molecular clouds \citep[GMCs;][]{1985sanders} which are the birthplaces of new stars.
Multiple surveys of extragalactic CO and and its main isotopologues (e.g. PAWS, PHANGS, VERTICO) have proven these lines to be good tracers of bulk molecular gas \citep{2012kennicutt}. The ratios of integrated intensities of these various CO emission lines allow to constrain physical and chemical conditions of the gas, such as excitation temperature and density \citep{2017Penaloza}. Particularly, CO isotopologue line ratios of the same $J$-transitions inform us about the opacity and abundance within the gas \citep{2015shirley}.

Studies of CO isotopologues include our Galaxy \citep[]{1990langer,1994wilson,2001Sawada,2010Yoda} and, beyond our Galaxy, on (ultra)-luminous infrared galaxies \citep[U/LIRGs;][]{2017bsliwa, 2019brown, 2021Pereira-Santaella}, starburst galaxies \citep[]{2004meier,2011costagliola,2013aladro, 2014Davis,2024hao}, galaxy centres \citep[]{2022davis,2023Teng} and early-type galaxies \citep[ETGs;][]{2010krips,2012Crocker,2013Davis}. In normal star-forming spiral galaxies, most studies have been conducted at kiloparsec-scale resolution \citep[]{2001paglione,2010krips,2011tan,2014Davis,2017jimenez,2018cormier,2022denbrok}. However, in recent years, there has been a growing interest in high-resolution mapping of CO and its isotopologues in these galaxies \citep[]{2013schinnerer,2013donovan,2016topal,2022egusa,2023denbrok,2023Koda,2024denbrok}.

One such effort is the SWAN survey \citep[Surveying the Whirlpool Galaxy at Arcseconds with NOEMA; PIs E. Schinnerer and F. Bigiel;][]{2023stuber,2025stuber}, which mapped several CO isotopologues and dense gas tracers at a spatial resolution of 3$\arcsec$ ($\approx$ 125 pc) in the nearby \citep[D = 8.58 Mpc;][]{2016mcquinn} face-on \citep[i = 22$^{\circ}$, PA = 173$^{\circ}$;][]{2014colombo_b} grand-design spiral, the Whirlpool galaxy (M51; NGC5194), that hosts a low-luminosity AGN \citep[]{1997Ho,2011dumas,2016querejeta}. What distinguishes SWAN from previous studies \citep{2008vila-vilaro, 2011tan,2018cormier,2022denbrok} is its ability to map these CO isotopologue lines across a significant portion of the M51 disk at a GMC-scale resolution ($\approx$ 100 pc). 

\begin{figure*}
\centering
\includegraphics[width=0.7\textwidth]{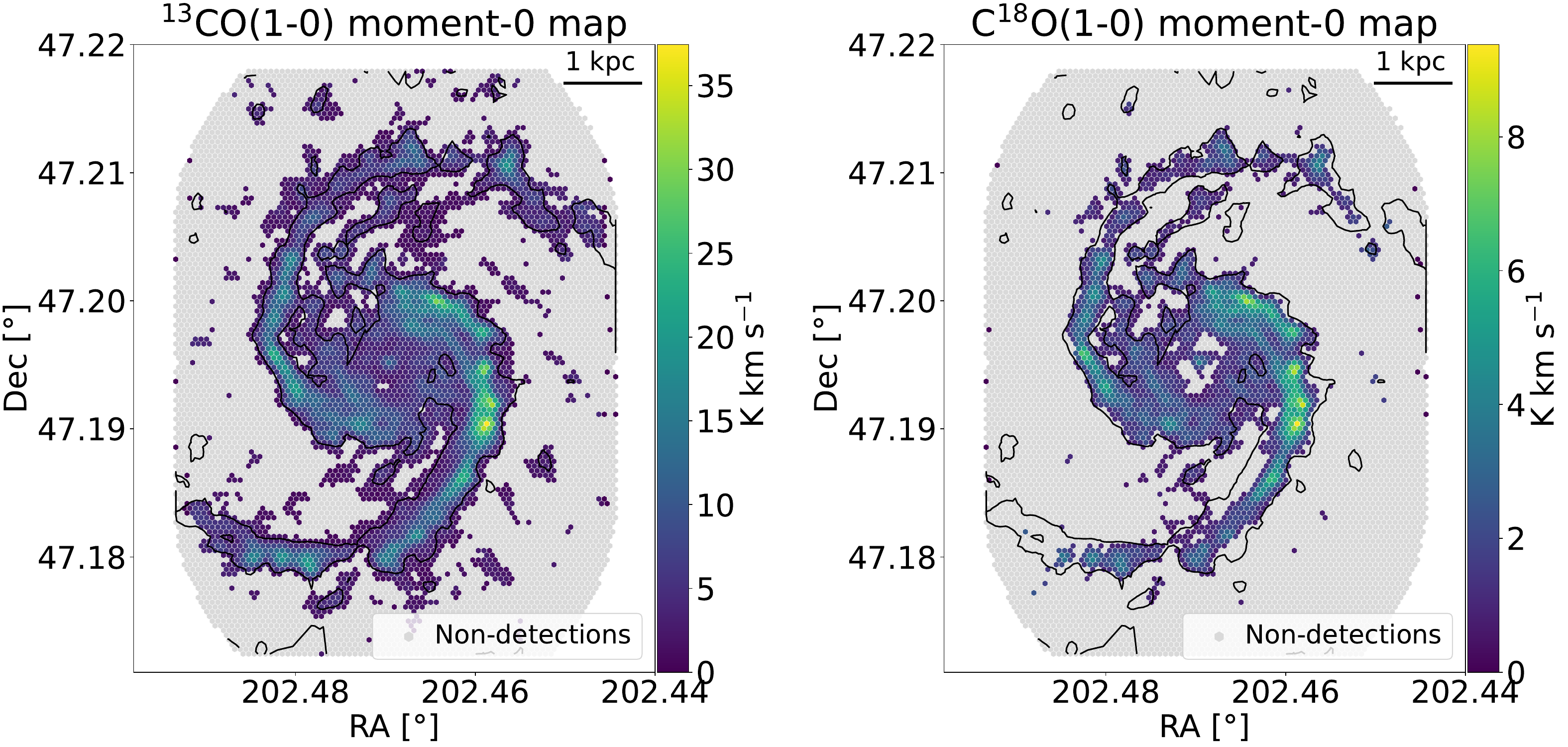}
\caption{The figure presents integrated intensity (moment-0) maps of the $^{13}$CO(1–0) (left) and C$^{18}$O(1–0) (right) emission lines. Gray points denote non-detections, i.e. sightlines with S/N $\leq$ 3, while coloured points indicate detections with S/N > 3. Overlaid contours correspond to the $^{12}$CO(1–0) emission \citep{2013pety} at the 30 K km s$^{-1}$ level, shown for reference.}
          \label{fig:lines}
\end{figure*}

Here we present an analysis of $^{13}$CO(1-0) and C$^{18}$O(1-0) line emission from the SWAN survey (see Fig. \ref{fig:lines}). The $^{13}$CO(1-0)/C$^{18}$O(1-0) line ratio can vary due to changes in opacity, underlying physical conditions, or isotopic abundances driven by mechanisms such as chemical fractionation, selective photodissociation, or selective nucleosynthesis. To understand these variations, we examine the ratio of integrated intensities of these lines ($\mathrm{{R}^{13}_{18}}$) across different environments in M51 and how it links to star formation rate (SFR) and surface density ($\Sigma_{\text{SFR}}$).

\begin{figure*}
\centering
\includegraphics[width=\textwidth]{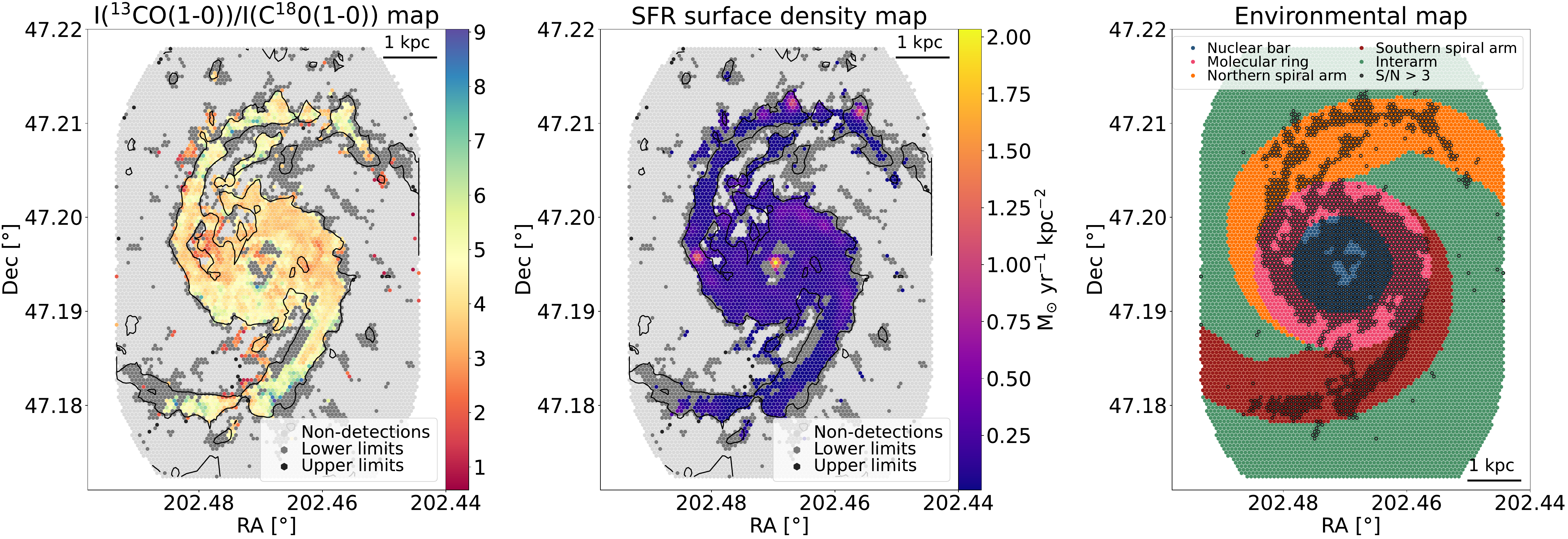}
\caption{The complete dataset is presented in this figure, while the version masked for AGN activity and used in the analysis is provided in Appendix \ref{AGN Acivity}. The left panel shows the $\mathrm{R}^{13}_{18}$ line ratio map, while the central panel presents the $\Sigma_{\text{SFR}}$ map, limited to regions where the line ratio is significantly measured. In these maps, light gray points indicate non-detections (S/N $\leq$ 3 in both lines), intermediate gray denotes lower limits (S/N $\leq$ 3 in C$^{18}$O) and dark gray indicates upper limits (S/N $\leq$ 3 in $^{13}$CO). Coloured points mark detections with S/N > 3 in both lines. The overlaid contours represent the 30 K km s$^{-1}$ level of $^{12}$CO(1–0) emission for reference. The right panel shows the PAWS environmental mask \citep{2014colombo} over the SWAN FoV, where different colours denote distinct environments: nuclear bar (blue), molecular ring (pink), northern spiral arm (orange), southern spiral arm (red), and interarm (green). Points with black outlines correspond to the significant detections shown in the other panels.}
          \label{fig:maps}
\end{figure*}

\section{Data}

The SWAN survey, an IRAM large program, (LP003; PIs: F. Bigiel, E. Schinnerer) used the NOrthern Extended Millimetre Array (NOEMA) and the IRAM 30-meter telescope to map 3-4 mm emission lines in M51 at a spatial resolution of 3$\arcsec$ ($\approx$ 125 pc) over the central 5 $\times$ 7 kpc$^2$. For a more detailed description of the observations and data reduction, see Appendix A in \cite{2023stuber} and \cite{2025stuber}. Additional data include Spitzer 24 $\mu$m maps from \citet{2011dumas} and the H$\alpha$ maps from \citet{2020kessler} combined using Equation 6 from \citet{2013leroy} to create a $\Sigma_{\text{SFR}}$ map (see middle panel in Fig. \ref{fig:maps}), as well as the $^{12}$CO(1-0) map from the PdBI Arcsecond Whirlpool Survey \citep[PAWS;][]{2013schinnerer}.  For our environmental analysis we use a simplified version of the PAWS environmental mask \citep{2014colombo} based on the underlying stellar potential. We do not account for dynamically distinct subregions within the spiral arms or interarm regions; instead, we consider only the following broader environments (see right panel in  Fig.\ref{fig:maps}): the nuclear bar, the molecular ring, the northern arm, the southern arm, and the interarm. For ease of comparison to lower resolution works, we also group these environments into two more general categories of the centre (defined as R < 1.3 kpc, containing the nuclear bar and molecular ring) and the disk (1.3 kpc < R < 5 kpc, containing both spiral arms and the interarm region). Using the PyStructure code \citep{2023neumann}, the datasets were convovled to the same spatial resolution of 3.05$\arcsec$ and spectral resolution of 10 km s$^{-1}$ for CO and its isotopologues before a half-beam-sized sampling was applied. Moment maps were created by integrating voxels where $^{12}$CO or HCN was detected, with the detected region defined using a 3D mask with thresholds of S/N > 4 expanded to S/N > 2.  Since the influence of the AGN on the CO isotopologues is not yet well understood, we mask the central region to avoid introducing potential biases into our analysis. A more detailed discussion of this decision is provided in Section \ref{Variations in Molecular Abundances}. Specifically, we exclude the inner 500 pc, which encompasses the AGN radio jets \citep{2016querejeta_b}. Although checks based on the ionization state of the gas—following the methodology of \citet{2019dagostino}—suggest that masking the inner 300 pc would suffice, we adopt a more conservative approach by extending the exclusion to 500 pc to ensure minimal contamination from possible AGN-related effects.

\section{Results}

\subsection{Ratio of medians}\label{Ratio of Medians}

In this paper, we focus on $\mathrm{{R}^{13}_{18}}$ as a way to trace variations of isotopic abundance and opacity in molecular clouds, particularly in relation to SF activity and galactic structure. Figure \ref{fig:maps} (left) displays a map of $\mathrm{{R}^{13}_{18}}$ across M51. For sightlines where both lines have S/N > 3, the 16th and 84th percentiles extend from 3.4 and 5.2, respectively. To understand how $\mathrm{{R}^{13}_{18}}$ varies between environments and how this local behaviour compares to the galaxy averaged behaviour, we calculate the ratio of medians both for the entire field-of-view (FoV) and for each specific environment using the following equation:
\begin{equation}
    \mathrm{\widetilde{{R}}^{13}_{18} = \widetilde{I}_{13} / \widetilde{I}_{18}}
    \label{eq:ratio_of_medians}
\end{equation}
where $\widetilde{\mathrm{I}}_{13}$ and $\widetilde{\mathrm{I}}_{18}$ represent the median integrated intensities of $^{13}$CO and C$^{18}$O lines, respectively.  We choose to use the ratio of median integrated intensities rather than the median of individual integrated intensity ratios, as this approach is less sensitive to C$^{18}$O detectability. When C$^{18}$O is faint or undetected, pixel-wise ratios can become artificially large, skewing the median of ratios. While applying a signal-to-noise cut could mitigate this, it would bias the analysis toward bright regions and misrepresent the full FoV. By using the ratio of medians without S/N cuts, we retain information from low-S/N sightlines while minimizing the impact of outliers. Specifically for the SWAN data, the relative difference\footnote{Calculated as $\frac{x -x_{\mathrm{ref}}}{x_{\mathrm{ref}}}$, where $x_{\mathrm{ref}}$ denotes the corresponding ratio of medians.} between the median of ratios and ratio of medians across the entire FoV is approximately 11\% of the ratio of medians value. When broken down by environment, the difference decreases to just 1–3\%, except in the southern spiral arm, where it increases to 21\%. For more information on the median of ratios and its uncertainties see Appendix \ref{Alternative Ratio Averaging Methods and Their Uncertainties}.

We apply Equation \ref{eq:ratio_of_medians} to the data, with the propagated statistical uncertainty calculated by combining $\widetilde{I}_{13}$ and $\widetilde{I}_{18}$ uncertainties using standard error propagation. The results are summarized in Table \ref{table:1}.  We now compare our results to values reported in the literature, noting that some of the referenced studies use different methodologies (e.g., median of ratios rather than ratio of medians). In such cases, the uncertainties provided in Table \ref{table:1} should be taken into account when assessing the level of agreement. The resulting $\mathrm{\widetilde{{R}}^{13}_{18}}$ across the FoV is 4.33 $\pm$ 0.05, aligning well with \citet{2022denbrok} who used the IRAM 30 m telescope to map $\approx$ 15 $\times$ 15 kpc$^2$ of M51 at kpc-scale resolution, reporting an average ratio of 4.35$_{-0.03}^{+0.02}$. For the centre, we find a $\mathrm{\widetilde{{R}}^{13}_{18}}$ of 3.98 $\pm$ 0.02, while for the disk, it is 4.8 $\pm$ 0.1, resulting in a relative difference of 20\% of the central value. The central value is close to the 3.6 $\pm$ 0.3 reported by \citet{2008vila-vilaro}, who found the ratio based on single-point observation of the inner 950 pc using the ARO Kitt Peak 12 m telescope. Similarly, \citet{2011tan}, who mapped M51's major axis with 13 pointings using the PMO 14 m telescope, found a ratio of 3.7 for the nucleus. However, their reported ratio of 2.6 for the disk differs from our findings. \citet{2017jimenez} observed the ratio ranging from 7 to 10 in the disks of nine nearby normal star-forming galaxies, which are significantly higher than our disk ratio. For galactic centres, they report ratios between 3.8 and 8.7, which are more consistent with our central value. In the Milky Way, ratios reported by \citet{1990langer} and \citet{2008wouterloot} span from 5 to 10, also exceeding the values we find for M51 in both the disk and the centre. When examining individual environments, the southern spiral arm exhibits a notably higher $\mathrm{\widetilde{{R}}^{13}_{18}}$, approximately 24\% above the FoV value, while the other environments show an average deviation of only 9\%. This deviation in the southern arm does not appear to be purely statistical, given its similar number of data points and scatter level compared to the northern arm, which does not exhibit a similar trend. Additionally, $\mathrm{{R}^{13}_{18}}$ is not the only line ratio exhibiting such behaviour; \citet{2023stuber} report an elevated N$_2$H$^{+}$(1-0)/HCN(1-0) ratio in the same area of the southern arm. This suggests that the variation may have a physical origin.

A potential explanation for these elevated line ratios could involve the difference in $\Sigma_{\rm {SFR}}$ between the arms, with the northern arm having a 25\% larger $\Sigma_{\rm {SFR}}$ than the southern one. \citet{2013meidt} and \citet{2016querejeta_b} have shown that the southern arm has a reduced star formation efficiency (SFE) due to strong streaming motions stabilizing the gas. The $\mathrm{\widetilde{{R}}^{13}_{18}}$ in the nuclear bar and the molecular ring, on the other hand, fall within the ranges typically observed for starbursts \citep[3 to 6;][]{1991sage,1995aalto} and on the higher end for ULIRGs \citep[0.2 to 4;][]{2017bsliwa,2019brown}.

\begin{table}
\caption{$\mathrm{\widetilde{{R}}^{13}_{18}}$ and median $\Sigma_{\rm {SFR}}$ by region}
\label{table:1}
\centering
\begin{tabular}{ccc}
\hline\hline
Region & $\mathrm{{\widetilde{R}}^{13}_{18}}$ & $\Sigma_{\rm {SFR}}$ [M$_{\odot}$ yr$^{-1}$ kpc$^{-2}$] \\
\hline\hline
Whole FoV &  4.33 $\pm$ 0.05 &  0.035 $\pm$ 0.002\\
Nuclear bar &  3.86 $\pm$ 0.03 &  0.140 $\pm$ 0.003 \\
Molecular ring & 3.88 $\pm$ 0.03 & 0.105 $\pm$ 0.005 \\
Northern spiral arm & 4.6 $\pm$ 0.1 & 0.035 $\pm$ 0.005 \\
Southern spiral arm & 5.4 $\pm$ 0.2 & 0.027 $\pm$ 0.002 \\ 
Centre &  3.98 $\pm$ 0.02 &  0.117 $\pm$ 0.004 \\
Disk & 4.8 $\pm$ 0.1 & 0.025 $\pm$ 0.002 \\
\hline
\end{tabular}
\tablefoot{The table presents the ratio of medians for $\mathrm{{R}^{13}_{18}}$ (middle column) with its propagated statistical uncertainty and the median $\Sigma_{\rm {SFR}}$ (right column) with its corresponding scatter, categorized by environment.}
\end{table}

\subsection{R$^{13}_{18}$ in different environments} \label{Local environmental conditions ratio correlations}

To investigate how $\mathrm{{R}^{13}_{18}}$ varies with local environmental conditions, we compare the ratio with galactocentric radius and $\Sigma_{\rm SFR}$. These parameters were selected because they effectively capture potential gradients in physical and chemical conditions. Specifically, SF can influence the chemical enrichment and excitation conditions of the gas, while factors such as gas-phase metallicity are known to vary with radius. To quantify the correlation between $\mathrm{{R}^{13}_{18}}$ and either galactocentric radius or $\Sigma_{\rm SFR}$, we use Kendall's $\tau$. This non-parametric method is well-suited for our data as it minimizes bias from outliers and accommodates the non-normal distribution of ratio values. We compute Kendall’s tau for both the entire FoV and each environment. We present two methods of doing this - one is by using only points with a S/N > 3 (shown as coloured points in  Fig.\ref{fig:env_plots_radius} and  Fig.\ref{fig:env_plots_sfr}) and the other is using spectrally stacked (see Appendix \ref{Spectral Stacking} for details) points with S/N > 3 (shown as hexagonal points in  Fig.\ref{fig:env_plots_radius} and  Fig.\ref{fig:env_plots_sfr}). The resulting correlation coefficients ($\tau$) are presented in Table \ref{table:2}.  For the stacking analysis, we tested using 6, 8, 10, and 12 bins to evaluate the stability of the correlation coefficients. We found that the choice of bin number had no significant effect on the overall trends. Specifically, using 10 bins resulted in relative differences of 11\%, 10\%, and 0.04\% compared to results obtained with 12, 8, and 6 bins, respectively. Based on this, we adopted 10 bins as a representative and stable choice. For reference, we categorize correlation coefficients as follows: values between 0.1 and 0.4 are considered low, 0.4 to 0.7 as moderate, and values above 0.7 as high.

\textbf{Galactocentric radius} The top panel in Fig.\ref{fig:env_plots_radius} illustrates how $\mathrm{{R}^{13}_{18}}$ varies with distance from the galactic centre. Sightlines where both $^{13}$CO(1-0) and C$^{18}$O(1-0) have S/N > 3 are shown as coloured points while the gray points seen in the background of the panel have one or both lines with S/N $\leq$ 3. If $^{13}$CO(1-0) is not detected, it indicates an upper limit, whereas if C$^{18}$O(1-0) is not detected, it indicates a lower limit. The scatter of all points is $\approx$ 0.4 dex. With $\tau$ = 0.22 (p-value $\ll$ 0.05), the FoV region shows a low correlation between the ratio and radius. When calculated from the stacked points, the correlation coefficient increases to $\tau$ = 0.71 (p-value = 0.06), suggesting a high correlation.

\begin{table}
\caption{Correlation coefficients between $\mathrm{{R}^{13}_{18}}$ and environment}
\label{table:2}
\centering
\begin{tabular}{ccc}
\hline\hline
 Regions & Pixels & Stacks \\
\hline\hline
Whole FoV & -0.18 ($\ll$ 0.05) & -0.86 ($\ll$ 0.05) \\
Nuclear bar &  -0.08 ($\ll$ 0.05) &  -0.51 (0.05) \\
Molecular ring & -0.11 ($\ll$ 0.05) & -0.72 ($\ll$ 0.05) \\
Northern spiral arm & -0.07 ($\ll$ 0.05) & -0.89 ($\ll$ 0.05) \\
Southern spiral arm & 0.12 ($\ll$ 0.05) & -0.21 (0.55) \\ 
\hline\hline
Whole FoV &  0.22 ($\ll$ 0.05) &  0.71 ($\ll$ 0.05)\\
\hline
\end{tabular}
\tablefoot{This table summarizes the Kendall's $\tau$ for $\mathrm{{R}^{13}_{18}}$ in relation to galactocentric radius across the FoV (bottom row) and with $\Sigma_{\rm SFR}$ across different environments (remaining rows). The coefficients are calculated for pixels with S/N > 3 (middle column) and for stacked points (right column). Bracketed numbers indicate the corresponding p-value.}
\end{table}

\textbf{SFR surface density} We additionally investigate $\mathrm{{R}^{13}_{18}}$ as function of $\Sigma_{\rm SFR}$ in Fig.\ref{fig:env_plots_sfr} and find a negative trend. However, the scatter at low $\Sigma_{\rm SFR}$ also increases as we approach the detection limits of the CO isotopologues and SFR. We observe a low negative correlation across the FoV with $\tau$ = -0.18 (p-value $\ll$ 0.05) and a stronger negative trend of $\tau$ = -0.86 (p-value $\ll$ 0.05) when using the stacked points. Examining individual environments, the point-based method yields low correlations with statistically significant p-values, expect for the nuclear bar and northern arm, which show no correlation. These correlations are negative in all regions except for the southern spiral arm. In contrast, the stacked method indicates high correlation coefficients with significant p-values for the molecular ring and northern spiral arm, while the nuclear bar and southern spiral arm exhibit low to moderate correlation coefficients with high p-values.

While there is significant scatter and the actual median ratio varies roughly by 68\% with radius in the inner 3.2 kpc and by 30\% with $\Sigma_{\rm SFR}$ for $\Sigma_{\rm SFR}$ > 0.02, formally we find a correlation between the ratio and galactocentric radius, as well as an anti-correlation with $\Sigma_{\rm SFR}$. We calculate these variations by determining the median ratio within each bin shown in Figs. \ref{fig:env_plots_radius} and \ref{fig:env_plots_sfr}, then dividing the difference between the maximum and minimum bin median ratios by the median ratio of the entire dataset. The observed trends are consistent with findings from \citet{2017jimenez} and \citet{2022denbrok} at $\approx$ kpc resolution. While we cannot precisely determine the strength of these correlations, we can establish some constraints on their values. The correlation coefficients from the stacked spectra are higher than those from the pixel-by-pixel analysis, primarily because stacking averages out noise and astrophysical scatter. The pixel-by-pixel method retains more local variability, leading to greater scatter in the correlations.
Comparing our correlation coefficients from stacked points to those of \citet{2022denbrok}, we find they report a slightly higher correlation coefficient of 0.8 (p = 0.083) for galactocentric radius, likely due to their extended detection range up to 5 kpc capturing higher ratio values. In the case of $\Sigma_{\rm SFR}$, they find a correlation of -0.67 (p = 0.33), which is lower than our value but has a higher p-value. While differences likely arise due to mismatched spatial scales, the trends align qualitatively.

\begin{figure*}[h!]
\centering
\includegraphics[width=\textwidth]{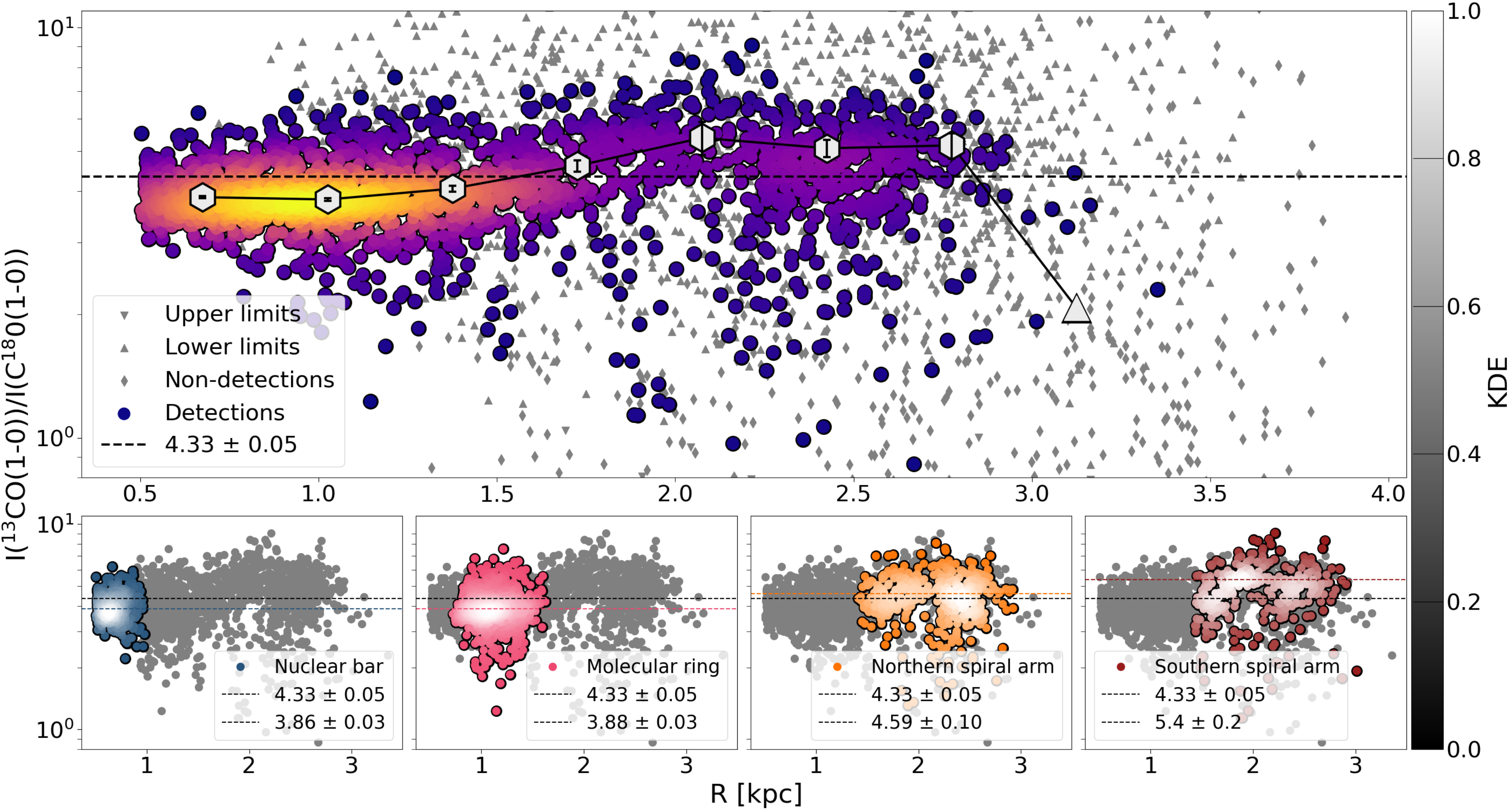}
\caption{The figure displays $\mathrm{{R}^{13}_{18}}$ plotted against galactocentric radius. In the top panel, coloured points represent sightlines where both emission lines have S/N > 3. Downward triangles denote lower limits, where the ratio has S/N $\leq$ 3 and C$^{18}$O(1-0) also has S/N $\leq$ 3. Upward triangles indicate upper limits, where the ratio has S/N $\leq$ 3 and $^{13}$CO(1-0) also has S/N $\leq$ 3. Diamonds represent non-detections, where both lines have either S/N $\leq$ 3 or S/N > 3, but result in the ratio having S/N $\leq$ 3. White hexagons correspond to points obtained via spectral stacking with the error bars corresponding to the propagated statistical uncertainties. The bottom panels highlight the points of each environment (see right map in  Fig. \ref{fig:maps}), with grey points matching the coloured ones in the top panel and the coloured points highlighting the respective environment. The black dashed line marks the $\widetilde{{\mathrm{R}}}^{13}_{18}$ for the FoV, while the coloured dashed line indicates the $\widetilde{{\mathrm{R}}}^{13}_{18}$ for the specific environment. The colour saturation for points in both the top and bottom panels reflects the kernel density estimate (KDE).}
          \label{fig:env_plots_radius}
\end{figure*}

\begin{figure*}[h!]
\centering
\includegraphics[width=\textwidth]{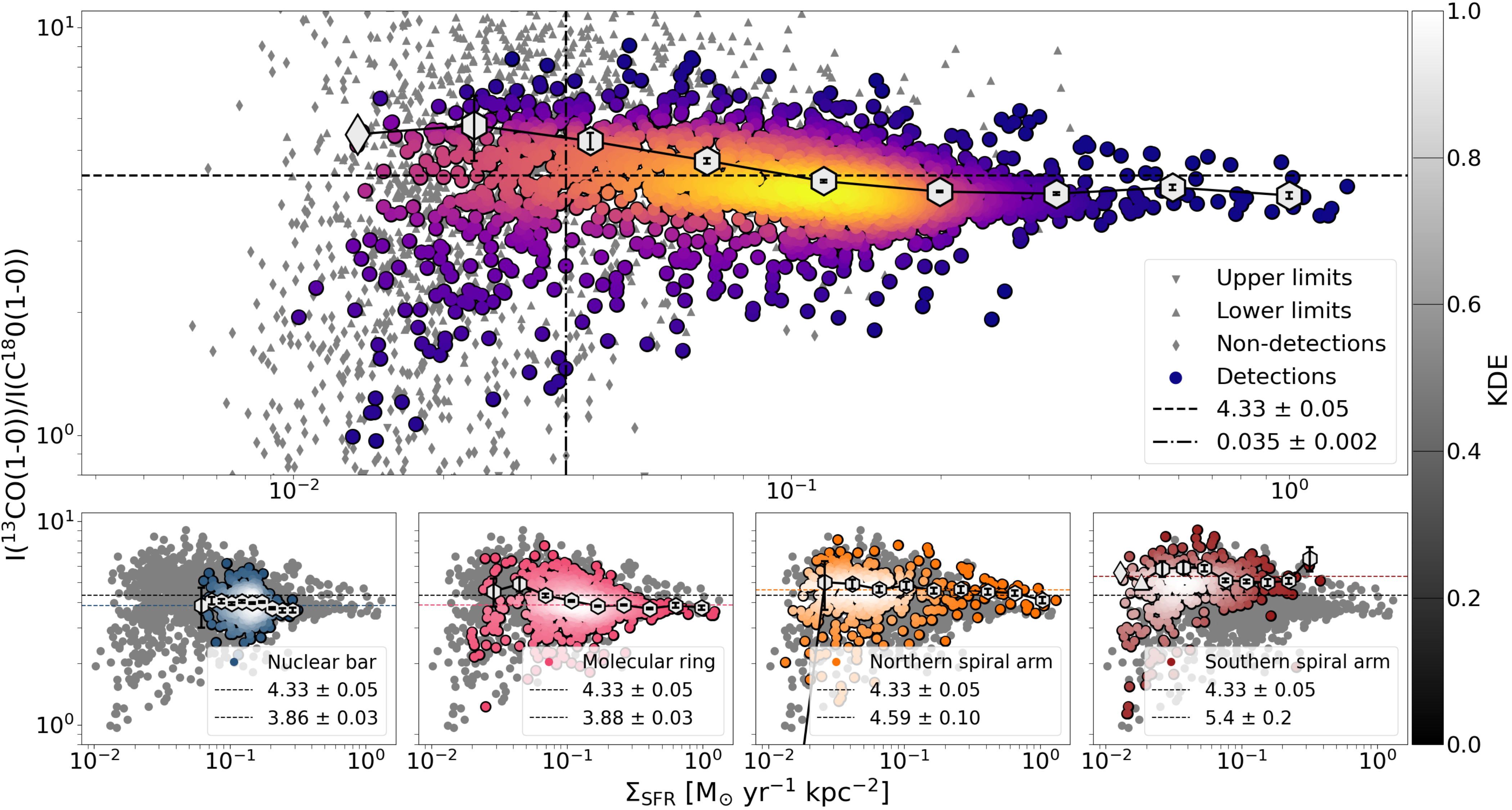}
\caption{The figure displays $\mathrm{{R}^{13}_{18}}$ plotted against SFR surface density. The description is analogous to Fig. \ref{fig:env_plots_radius}, with an added black dashed line indicating the FoV's median $\Sigma_{\text{SFR}}$. The black line extending in the northern spiral arm panel arises from a low-value non-detection point.}
          \label{fig:env_plots_sfr}
\end{figure*}

\section{Discussion}

The observed trends with $\mathrm{{R}^{13}_{18}}$ may be explained by abundance-driven variations resulting from processes such as selective nucleosynthesis, chemical fractionation, and selective photodissociation. Alternatively, changes in the opacities of both species with their abundances fixed or variations in line excitation could also account for these observed patterns. In this Section, we will explore the likelihood of those explanations for our observed trends.

\subsection{Variations in line excitation}
Previous studies conducted at kpc-scales \citep{2022denbrok}, along with more recent investigations \citep{2024denbrok} utilizing CO multi-line datasets from SMA, SWAN and PAWS data at a spatial resolution of 4$\arcsec$ ($\approx$170 pc), both conclude that, on global scales, variations in line excitation do not significantly affect the line ratios.

\subsection{Variations in molecular abundances}\label{Variations in Molecular Abundances}

\textbf{Chemical fractionation} In cold regions, the abundances of $^{13}$CO and C$^{18}$O molecules can be affected by chemical fractionation \citep[]{1976watson,1980smith,1998keene,2017romano} which enhances the production of $^{13}$CO at the expense of $^{12}$CO via the exothermic reaction:
\[
	\mathrm{^{13}C^{+} + \, ^{12}CO \longrightarrow \, ^{12}C^{+} + \, ^{13}CO + \Delta E}
\]
where $\Delta$E represents the energy change associated with the reaction.  \citep{1984langer} note that the fractionation reaction for C$^{18}$O,
\[
\mathrm{^{18}O + \, C^{16}O \longrightarrow \, ^{16}O + \, C^{18}O + \Delta E},
\]
is unlikely to occur since it has a small rate coefficient at 300 K and a substantial activation energy barrier. This makes the reaction inefficient at the low temperatures where $^{13}$CO fractionation is observed. Therefore, in a simple scenario where $\Sigma_{\text{SFR}}$ traces temperature, we would expect a high $\mathrm{{R}^{13}_{18}}$ in cold regions with low $\Sigma_{\text{SFR}}$, as shown in Fig. \ref{fig:env_plots_radius}. \citet{2014szucs} have performed hydrodynamical simulations of molecular clouds and found that chemical fractionation can decrease $^{12}$CO(1-0)/$^{13}$CO(1-0) $(\mathrm{{R}^{12}_{13}}$) by a factor of 2-3 at low temperatures and optical depths. We would then expect $\mathrm{{R}^{12}_{13}}$ to increase with $\Sigma_{\text{SFR}}$, but upon examining  Fig.\ref{fig:sfr1} we find that this is not the case on a galaxy-wide scale. Therefore, we conclude that this chemical reaction is unlikely to be the driver of the observed trends for R$^{13}_{18}$. These results are consistent with those of \citet{2018cormier}, who found a mild anti-correlation between $\mathrm{{R}^{12}_{13}}$ and both dust temperature and $\Sigma_{\text{SFR}}$ in M51.

\textbf{Selective photodissociation} The UV radiation from O and B stars, traced by regions of high SFR surface density, preferentially photodissociates $^{13}$CO and C$^{18}$O compared to $^{12}$CO molecules due to their poorer self-shielding capabilities resulting from lower abundances \citep[]{1988vandishoeck,2005lyons}). Since $^{13}$CO is slightly more abundant than C$^{18}$O, it is less susceptible to photodissociation, leading to a higher R$^{13}_{18}$ ratio as $\Sigma_{\text{SFR}}$ increases. This process should also cause $\mathrm{{R}^{12}_{13}}$ to increase with $\Sigma_{\text{SFR}}$. However, Fig. \ref{fig:sfr1} reveals that $\mathrm{{R}^{12}_{13}}$ decreases with $\Sigma_{\text{SFR}}$ in M51, suggesting that selective photodissociation is also an unlikely contributor to the observed trends.

\textbf{Selective nucleosynthesis} Stars of varying masses and ages enrich the ISM with different CO isotopologues \citep[]{1993henkel,1992casoli}, making these molecules valuable indicators of a galaxy's evolutionary stage. Specifically, $^{13}$C is synthesized during the helium-burning phase of intermediate-mass stars, while $^{18}$O and $^{12}$C are produced in supernova explosions of high-mass stars. If selective nucleosynthesis were driving the observed trends, we would expect increased C$^{18}$O production in regions of high $\Sigma_{\text{SFR}}$, thereby lowering R$^{13}_{18}$. Furthermore, by observing nine interstellar clouds in the Galaxy, \citet{1990langer} and \citet{2005milam} found that $^{12}$C/$^{13}$C and $^{16}$O/$^{18}$O ratios increase from the galactic centre, which is attributed to the inside-out formation of galaxies. This is consistent with the trends we observe for the plots in Appendix A and B and Fig. \ref{fig:env_plots_sfr}.

\subsection{Variations in opacity}
Another potential explanation for the observed trends is opacity variations of the lines across the FoV. This was investigated in \citet{2024denbrok}, where the optical depth of $^{13}$CO(1-0) is derived under both LTE and non-LTE conditions.
The $^{13}$CO optical depth map presented in the study shows an increase in opacity towards the galaxy’s centre ($\approx$ 0.1 - 0.6), leading to a lower emissivity of $^{13}$CO gas in the central region. This reduction in $^{13}$CO emission decreases R$^{13}_{18}$, potentially explaining the radial trend we observe.
To quantify this effect, we use the solution to the radiative transfer equation. Assuming no background intensity, similar excitation conditions and beam filling factors between the two molecular lines, we arrive at the following expression: $I_1/I_2 = (1 - e^{-\tau_1})/(1 - e^{-\tau_2})$, where $\tau_i$ represents the optical depth and I$_i$ the integrated intensity of each line. To quantify the impact of optical depth, we assume a fixed [$^{13}$CO]/[C$^{18}$O] abundance ratio across the field of view. In the disk, $^{13}$CO is optically thin according to \citet{2024denbrok}, who report $\tau^{\mathrm{disk}}_{13} \approx 0.1$. Under these assumptions, the above equation simplifies to $I_1/I_2 \approx \tau_1/\tau_2 $ and since $\tau_1/\tau_2 = X_1/X_2$ we obtain $ X_{13}/X_{18} \approx 4.8 \, \pm \, 0.1 $. In the centre, $^{13}$CO becomes partially optically thick, with an average $ \tau^{\mathrm{centre}}_{13} \approx $ 0.6. Using $ \tau^{\mathrm{centre}}_{18} = (X_{18}/X_{13}) \times \tau^{\mathrm{centre}}_{13}$, we obtain $\tau^{\mathrm{centre}}_{18}$ = 0.0208 $\pm$ 0.0005. Substituting these optical depths back into the above equation yields $ I_{13} / I_{18} \approx 3.8387 \, \pm \, 0.0002 $ for the centre, which is slightly lower than the 3.98 $\pm$ 0.02 from observations. This suggests that changes in optical depth could contribute to the observed R$^{13}_{18}$ trends. We extended this analysis to the $^{12}$CO/$^{13}$CO ratio, assuming $^{12}$CO is optically thick across the entire FoV. For the disk, we calculate $ I_{12} / I_{13} = 1/\tau^{\mathrm{disk}}_{13}$ = 10, while for the centre we get $ I_{12} / I_{13} = 1/(1-e^{-\tau^{\mathrm{centre}}_{13}})$ = 2.21. From observations, we find $ I_{12} / I_{13} = 10.89 \, \pm \, 0.07 $ for the disk and $ I_{12} / I_{13} = 6.96 \, \pm \, 0.02 $ for the centre. The calculated disk value aligns well with the observed ratio, supporting the assumption of optically thin $^{13}$CO and thick $^{12}$CO in this region. However, the observed central ratio significantly exceeds the predicted value, suggesting the influence of additional processes driving the ratio higher in the central region.  \cite{2001paglione} demonstrated that higher velocity dispersion, elevated gas kinetic temperature or lower gas column density can reduce the optical depth of $^{12}$CO, thereby enhancing its intensity. \cite{2024teng} similarly find that dynamical effects can lower $^{12}$CO optical depth. Physical processes such as turbulence or the presence of bars in galaxy centres can drive these conditions, potentially leading to an increase in R$^{12}_{13}$ \citep{2004paglione,2009aisrael, 2009bisrael, 2018cormier}.

\subsection{Outliers from average trends}
\begin{figure}
\centering
\includegraphics[width=\linewidth]{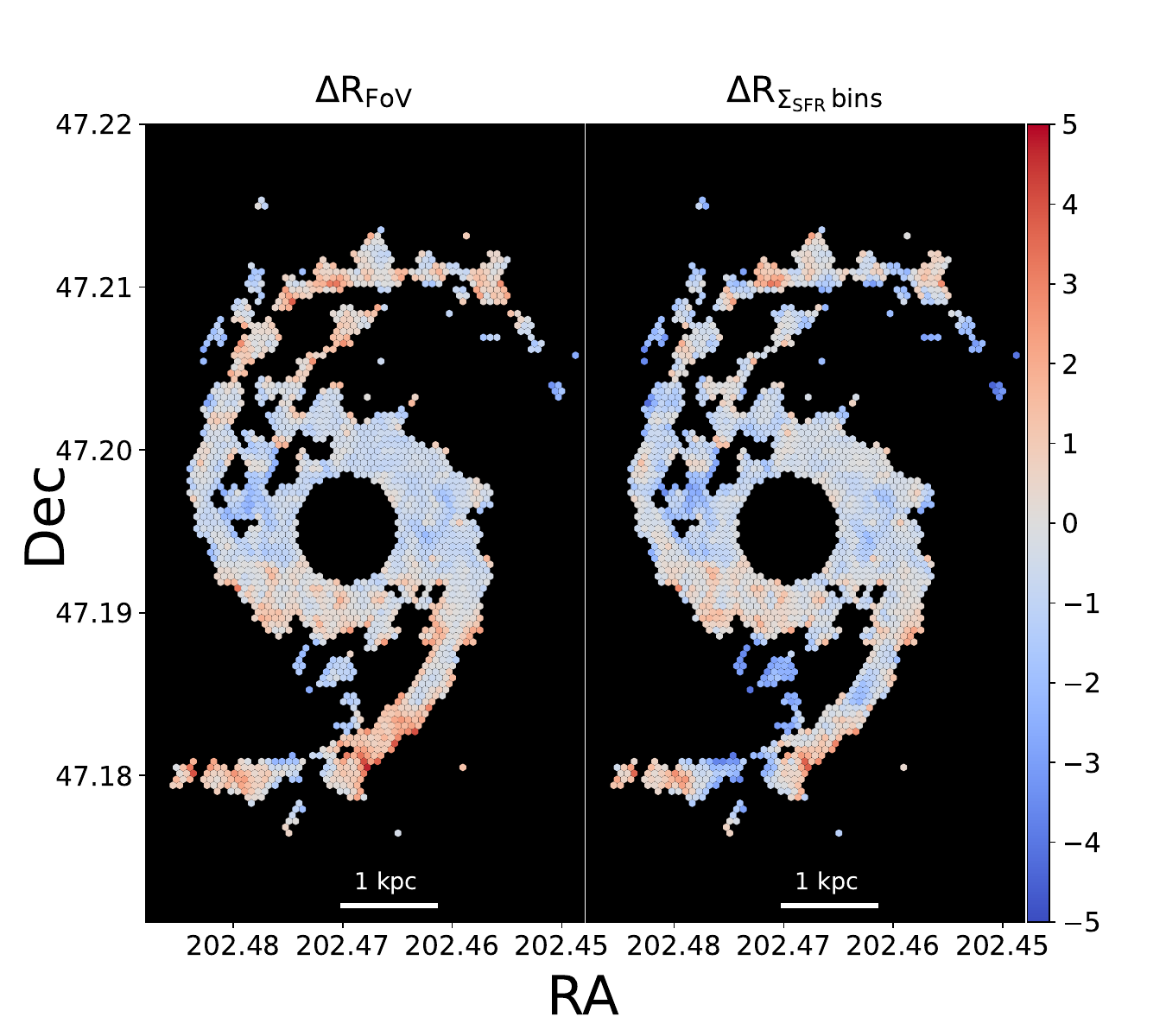}
\caption{The figure display deviations of significant (S/N $>$ 3) points from the average trends. The left panel illustrate the deviation of significant data points from the ratio of medians for the entire FoV while the right panels show how these significant points deviate from the stacked average within the corresponding $\Sigma_{\mathrm{SFR}}$ bin for the full FoV.}
          \label{fig:delta_maps}
\end{figure}

 To evaluate deviations from the average trends, we calculate the offset of statistically significant sightlines from both the FoV-wide ratio of medians as well as from the stacked values (gray hexagons in Figs. \ref{fig:env_plots_radius} and \ref{fig:env_plots_sfr}), which represent an unbiased average of the ratio as a function of galactocentric radius and $\Sigma_{\text{SFR}}$. We refer to these offsets from average values as $\Delta R$. For a sightline $i$, the $\Delta R$ from the FoV-wide ratio of medians is determined as
\begin{equation}
    \Delta R_{FoV, \, i} = R^{13}_{18, \, i} - \widetilde{R}^{13}_{18, \, FOV}
\end{equation}
where $R^{13}_{18, \, i}$ is the ratio of sightline $i$ and $\widetilde{R}^{13}_{18, \, FoV}$ is the FoV-wide ratio of medians. The variation in the ratio for sightline $j$ relative to the spectrally stacked value in the corresponding $\Sigma_{\text{SFR}}$ bin is calculated as
\begin{equation}
    \Delta R_{\Sigma_{\text{SFR}},\, j} = R^{13}_{18, \, j} - <R^{13}_{18}>_{stacked}
\end{equation}
where $R^{13}_{18, \, j}$ is the ratio for sightline $j$ and $<R^{13}_{18}>_{stacked}$ is the spectrally stacked ratio of all sightlines in the $\Sigma_{\text{SFR}}$ bin which contains sightline $j$. The scatter around this trend is considerable, with a typical deviation of $\approx$1 and a maximum deviation of $\approx$ 4.5, indicating that some sightlines reach up to twice the average ratio. The spatial distribution of these “outliers” is shown in Fig. \ref{fig:delta_maps}: the left panel shows deviations from the FoV-wide ratio of medians, while the right panel shows deviations from the $\Sigma_{\text{SFR}}$ stacked value in a corresponding bin. These maps clearly highlight regions with elevated or suppressed ratios across the FoV. The high $\Delta$ values in the northern arm align with regions of high $\Sigma_{\text{SFR}}$, suggesting a link to star formation activity, while a notable depletion is visible in the southern arm. As already seen in the ratio map in Fig. 1, the northwest side of the molecular ring shows a noticeably different characteristic ratio compared to the southeast side. This asymmetry remains evident even after controlling for $\Sigma_{\text{SFR}}$, indicating that this feature cannot be explained by star formation alone. Instead, it likely arises from other physical processes such as gas outflows or dynamical disturbances.

We conclude that abundance variations from selective nucleosynthesis and/or changes in the opacity of the gas are the most likely drivers of the observed trends with radius and $\Sigma_{\text{SFR}}$. While our study benefits from high spatial resolution and that of multiple lines compared to earlier work, our limited spatial coverage poses a restriction. Expanding our observations to encompass larger radii and lower SFR regions could better constrain the strength of the correlations we find. Additionally, our current approach to calculating $\Sigma_{\text{SFR}}$ relies on a combination of 24 $\mu$m and H$\alpha$ emission, which, as reported by \citet{2024calzetti}, can vary depending on the mean age of the stellar populations contributing to dust heating.

\section{Summary}
We present $^{13}$CO(1-0) and C$^{18}$O(1-0) observations of the inner 5 x 7 kpc$^2$ of M51 from the SWAN survey. We analyse the ratio of these lines against both the galactocentric radius and the SFR surface density across the SWAN FoV, as well as within distinct environments encompassed by the FoV.

Our key findings are as follows:

\begin{enumerate}
    \item The ratio of medians for the FoV is $\widetilde{{\mathrm{R}}}^{13}_{18}$ = 4.33 $\pm$ 0.05.
    \item The ratio shows a low positive correlation with galactocentric radius.
    \item The ratio shows a low negative correlation with $\Sigma_{\text{SFR}}$, particularly within the molecular ring and northern spiral arm.
    \item There is a notable difference in behaviour between the northern and southern spiral arms as expected from previous studies of the spiral structure in M51.
    \item The ratio trends are most likely driven by a combination of selective nucleosynthesis and optical depth effects.
\end{enumerate}

M51 provides valuable insights into the factors influencing CO isotopologues within a grand design spiral galaxy. It is evident that SF and environment play a role in this. Extension of such multi-line, high-resolution mapping of other galaxies could potentially reveal additional environmental dependencies across a wider range of SFR and radius values.

\begin{acknowledgements}
      This work was carried out as part of the PHANGS collaboration. This work is based on data obtained by PIs E. Schinnerer and F. Bigiel with the IRAM-30 m telescope and NOEMA observatory under project ID M19AA. This research was conducted with the support of the International Max Planck Research School (IMPRS) for Astronomy and Astrophysics at the universities of Bonn and Cologne. JEMD  acknowledges support from project UNAM DGAPA-PAPIIT IG 101025, Mexico. ES acknowledges funding from the European Research Council (ERC) under the European Union’s Horizon 2020 research and innovation programme (grant agreement No. 694343). SKS acknowledges financial support from the German Research Foundation (DFG) via Sino-German research grant SCHI 536/11-1. JPe acknowledges support by the French Agence Nationale de la Recherche through the DAOISM grant ANR-21-CE31-0010 and by the Programme National ``Physique et Chimie du Milieu Interstellaire'' (PCMI) of CNRS/INSU with INC/INP, co-funded by CEA and CNES. MJJD, AU and MQ acknowledge support from the Spanish grant PID2022-138560NB-I00, funded by MCIN/AEI/10.13039/501100011033/FEDER, EU. JdB acknowledges support from the Smithsonian Institution as a Submillimeter Array (SMA) Fellow. JEMD gratefully acknowledges funding from the Deutsche Forschungsgemeinschaft (DFG, German Research Foundation) in the form of an Emmy Noether Research Group (grant number KR4598/2-1, PI Kreckel) and the European Research Council’s starting grant ERC StG-101077573 (”ISM-METALS”). HAP acknowledges support from the National Science and Technology Council of Taiwan under grants 110-2112-M-032-020-MY3 and 113-2112-M-032 -014 -MY3. TAD acknowledges support from the UK Science and Technology Facilities Council through grant ST/W000830/1.
\end{acknowledgements}

\bibliographystyle{aa}
\bibliography{sample}

\begin{appendix}

\section{Spectral stacking}\label{Spectral Stacking}

Spectral stacking is a method used to improve the signal-to-noise ratio of faint emission lines, such as C$^{18}$O(1-0) in our case, by aligning and averaging spectra from different regions. To achieve this, we use the PyStacker code \citep{2023neumann_pystacker} on the dataset. The code bins the spectra based on a specific quantity, such as galactocentric radius or star formation rate surface density, and then averages them to improve the overall S/N. A critical step in this process is aligning the spectra to ensure there is no velocity offset between them. This is achieved by shifting each spectrum to a common velocity frame, typically defined by a high-S/N reference line like $^{12}$CO(1–0) or HI 21 cm. In our case, the $^{12}$CO(1–0) line was used as a prior to define the velocity field for the stacking. This velocity alignment ensures that the emission lines from different regions of the galaxy are centred at 0 km/s. Once aligned, the spectra are averaged by summing the shifted spectra and dividing by the number of spectra in the bin.
In Figs. \ref{fig:env_plots_radius}-\ref{fig:env_plots_sfr}, the stacked points were created by dividing the stacked $^{13}$CO(1-0) stacks by the C$^{18}$O(1-0) ones. If both stacked lines had an S/N > 3, the point was considered a detection and marked as a hexagonal point on the plot. If the ratio error (computed by propagating the statistical uncertainty of each line) was lower than or equal to 3 and the line in the denominator (C$^{18}$O) had S/N $\leq$ 3, the point was marked with a upward triangle and indicated a lower limit. If the line in the numerator ($^{13}$CO) had S/N $\leq$ 3, it was marked with a downward triangle and indicated an upper limit. Lines where both the ratio error and individual S/N were lower than or equal to 3 were considered non-detections and marked with diamond points. In our correlation coefficient calculations using Kendall’s tau, non-detected and limit stacked points were excluded. These steps help ensure that the stacked spectra accurately represent the underlying emission, while excluding unreliable data points. 

\section{Alternative line ratios and their implications}\label{Alternative Line Ratios and Their Implications}

Analogous to our analysis of the $^{13}$CO(1-0) to C$^{18}$O(1-0) ratio, we also examine the $^{12}$CO(1-0) to $^{13}$CO(1-0) and $^{12}$CO(1-0) to C$^{18}$O(1-0) ratios. Among these, the $^{12}$CO/$^{13}$CO ratio offers the most extensive coverage of the galaxy, as both $^{12}$CO(1-0) and $^{13}$CO(1-0) are detected with high significance. The ratios of medians for these alternative line ratios are summarized in Table \ref{table:alt_ratios}.
Across the entire FoV, $^{12}$CO(1-0) is typically $\approx$ 9 times brighter than $^{13}$CO(1-0) and $\approx$ 40 times brighter than C$^{18}$O(1-0). Consistent with the behaviour of the $^{13}$CO/C$^{18}$O ratio, the ratio of medians in the central region is lower than in the disk, with a relative difference of 56\% for $^{12}$CO/$^{13}$CO and 88\% for $^{12}$CO/C$^{18}$O. Additionally, variations between spiral arms are also evident for these ratios: the northern arm exhibits a 16\% lower ratio of medians for $^{12}$CO/$^{13}$CO and a 28\% lower value for $^{12}$CO/C$^{18}$O compared to the southern arm.
We applied the same methods described in Section \ref{Local environmental conditions ratio correlations} to calculate Kendall's $\tau$ correlation coefficients between the ratios and both galactocentric radius and $\Sigma_{\rm SFR}$. For the radius relations, we observe moderate positive correlations when using the pixel-by-pixel method and a strong positive correlation when using the stacked method. For the pixel-by-pixel analysis across the entire FoV and each environment, we observe a low anti-correlation between both ratios and $\Sigma_{\rm SFR}$. When using stacked data, $^{12}$CO/$^{13}$CO ratio shows a low anti-correlation, while the $^{12}$CO/C$^{18}$O ratio exhibits a moderate anti-correlation for the FoV, but the p-values for these coefficients are much higher than the 0.05 threshold. When looking at individual environments, we find that the nuclear bar remains uncorrelated while the molecular ring has a high anti-correlation both for the main and the alternative ratios. Interestingly, the southern spiral arm shows a strong anti-correlation with $\Sigma_{\rm SFR}$ for these ratios, in contrast to the $^{13}$CO/C$^{18}$O ratio, where no such correlation is found.

\begin{table}
\caption{Ratio of medians for $^{12}$CO(1-0)/$^{13}$CO(1-0) and $^{12}$CO(1-0)/C$^{18}$O(1-0) by region}
\label{table:alt_ratios}
\centering
\begin{tabular}{ccc}
\hline\hline
Region & $\mathrm{{\widetilde{R}}^{12}_{13}}$ & $\mathrm{{\widetilde{R}}^{12}_{18}}$  \\
\hline\hline
Whole FoV &  9.63 $\pm$ 0.04 &  41.8 $\pm$ 0.5\\
Nuclear bar &  6.24 $\pm$ 0.03 &  24.1 $\pm$ 0.2 \\
Molecular ring & 7.76 $\pm$ 0.03 & 30.1 $\pm$ 0.2 \\
Northern spiral arm & 9.36 $\pm$ 0.06 & 43.0 $\pm$ 0.9 \\
Southern spiral arm & 11.11 $\pm$ 0.08 & 60 $\pm$ 2 \\ 
Centre &  6.96 $\pm$ 0.02 &  27.7 $\pm$ 0.2 \\
Disk & 10.89 $\pm$ 0.07 & 52 $\pm$ 1 \\
\hline
\end{tabular}
\tablefoot{The table presents the ratio of medians for $^{12}$CO(1-0)/$^{13}$CO(1-0) (middle column) and $^{12}$CO(1-0)/C$^{18}$O(1-0) (right column), along with their propagated statistical uncertainties, categorized by environment.}
\end{table}

\begin{table}
\caption{Kendall's rank correlation coefficient between $^{12}$CO(1-0)/$^{13}$CO(1-0) ratio and $\Sigma_{\rm SFR}$}
\label{table:3}
\centering
\begin{tabular}{ccc}
\hline\hline
Region & Pixels & Stacks \\
\hline\hline
Whole FoV & -0.35 ($\ll$ 0.05) &  -0.79 ($\ll$0.05) \\
Nuclear bar & -0.14 ($\ll$ 0.05) &  -0.33 (0.22) \\
Molecular ring & -0.19 ($\ll$ 0.05) & -0.78 ($\ll$ 0.05) \\
Northern spiral arm & -0.26 ($\ll$ 0.05) & -0.5 (0.08) \\
Southern spiral arm & -0.33 ($\ll$ 0.05) & -0.89 ($\ll$ 0.05) \\ 
\hline
\end{tabular}
\tablefoot{The description is the same as Table \ref{table:1}.}
\end{table}

\begin{table}
\caption{Kendall's rank correlation coefficient between $^{12}$CO(1-0)/C$^{18}$O(1-0) ratio and $\Sigma_{\rm SFR}$}
\label{table:3}
\centering
\begin{tabular}{ccc}
\hline\hline
Region & Pixels & Stacks \\
\hline\hline
Whole FoV & -0.27 ($\ll$ 0.05) &  -0.79 (0.18) \\
Nuclear bar &  -0.17 ($\ll$ 0.05) &  -0.38 (0.16) \\
Molecular ring & -0.12 ($\ll$ 0.05) & -0.83 ($\ll$ 0.05) \\
Northern spiral arm & -0.18 ($\ll$ 0.05) & -0.67 ($\ll$ 0.05) \\
Southern spiral arm & -0.13 ($\ll$ 0.05) & -0.71 ($\ll$ 0.05) \\ 
\hline
\end{tabular}
\tablefoot{The description is the same as Table \ref{table:1}.}
\end{table}

\begin{figure*}[h!]
\centering
\includegraphics[width=\textwidth]{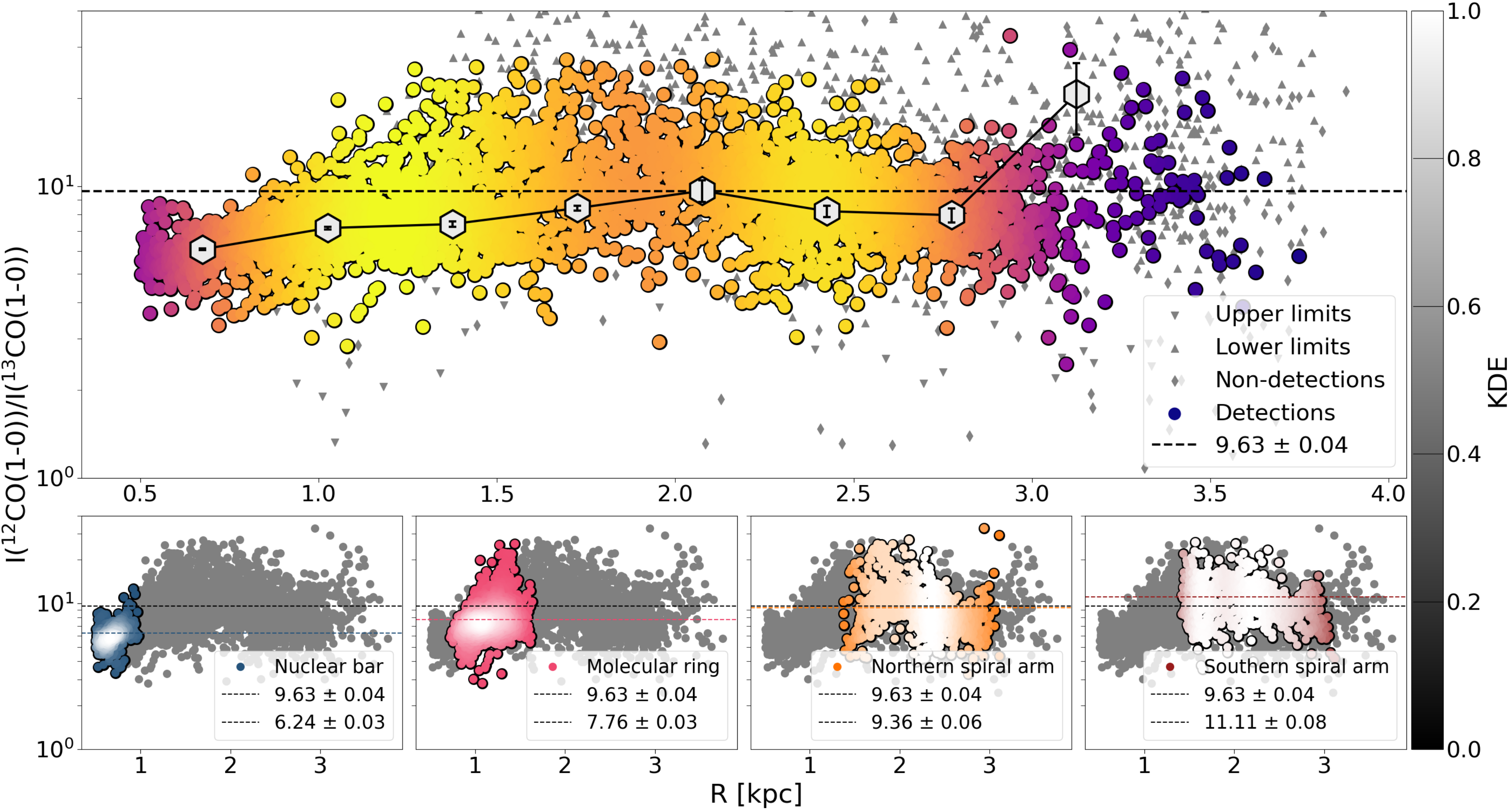}
\caption{The figure displays the $^{12}$CO(1-0) to $^{13}$CO(1-0) line ratio plotted against the galactocentric radius. The description is analogous to that in  Fig.\ref{fig:env_plots_radius}.}
\vspace{5mm}
\label{fig:radius1}
\end{figure*}

\begin{figure*}[h!]
\centering
\includegraphics[width=\textwidth]{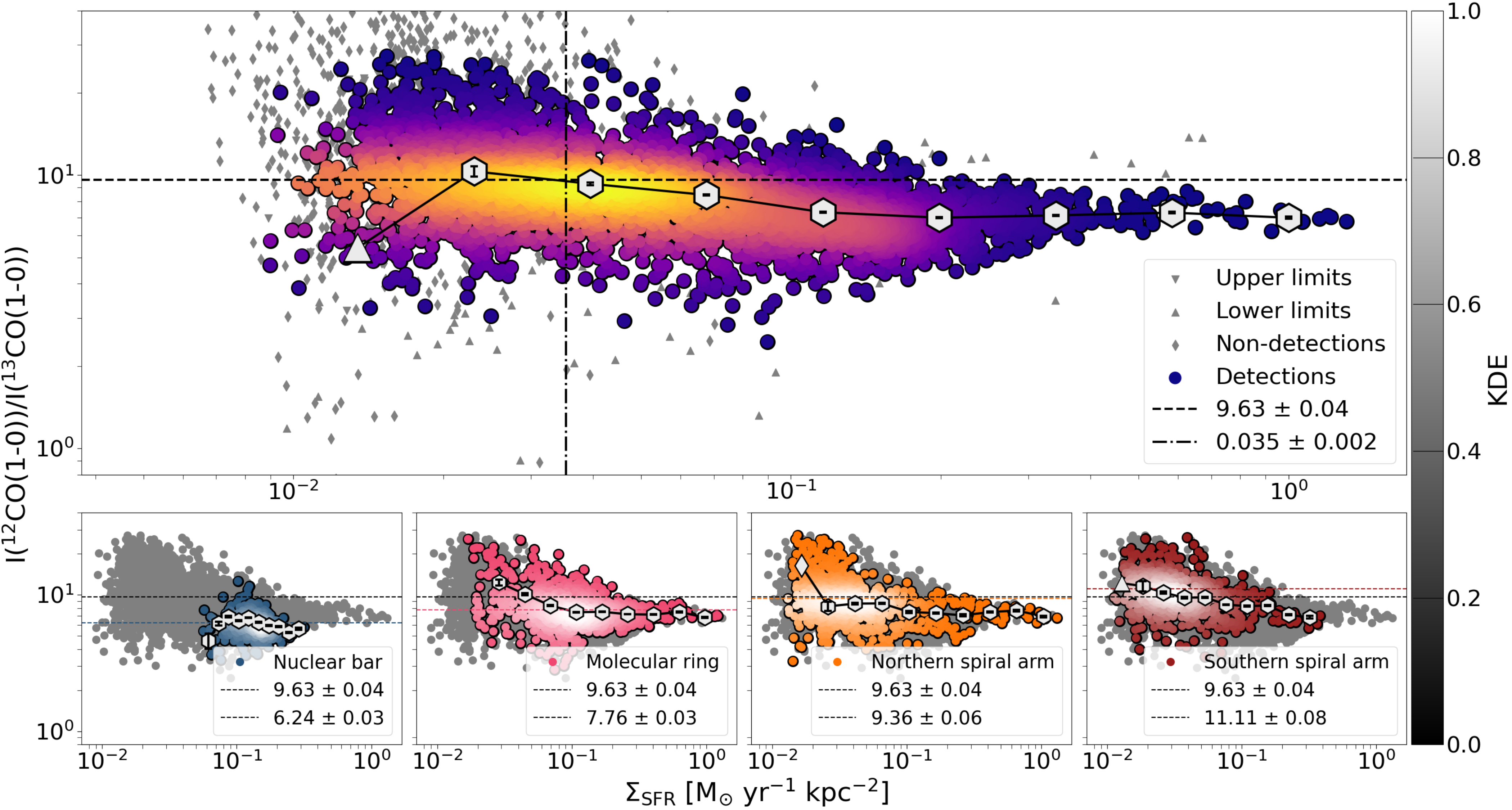}
\caption{The figure displays the $^{12}$CO(1-0) to $^{13}$CO(1-0) line ratio plotted against the SFR surface density. The description is analogous to that in  Fig.\ref{fig:env_plots_sfr}.}
\vspace{7mm}
\label{fig:sfr1}
\end{figure*}

\begin{figure*}[h!]
\centering
\includegraphics[width=\textwidth]{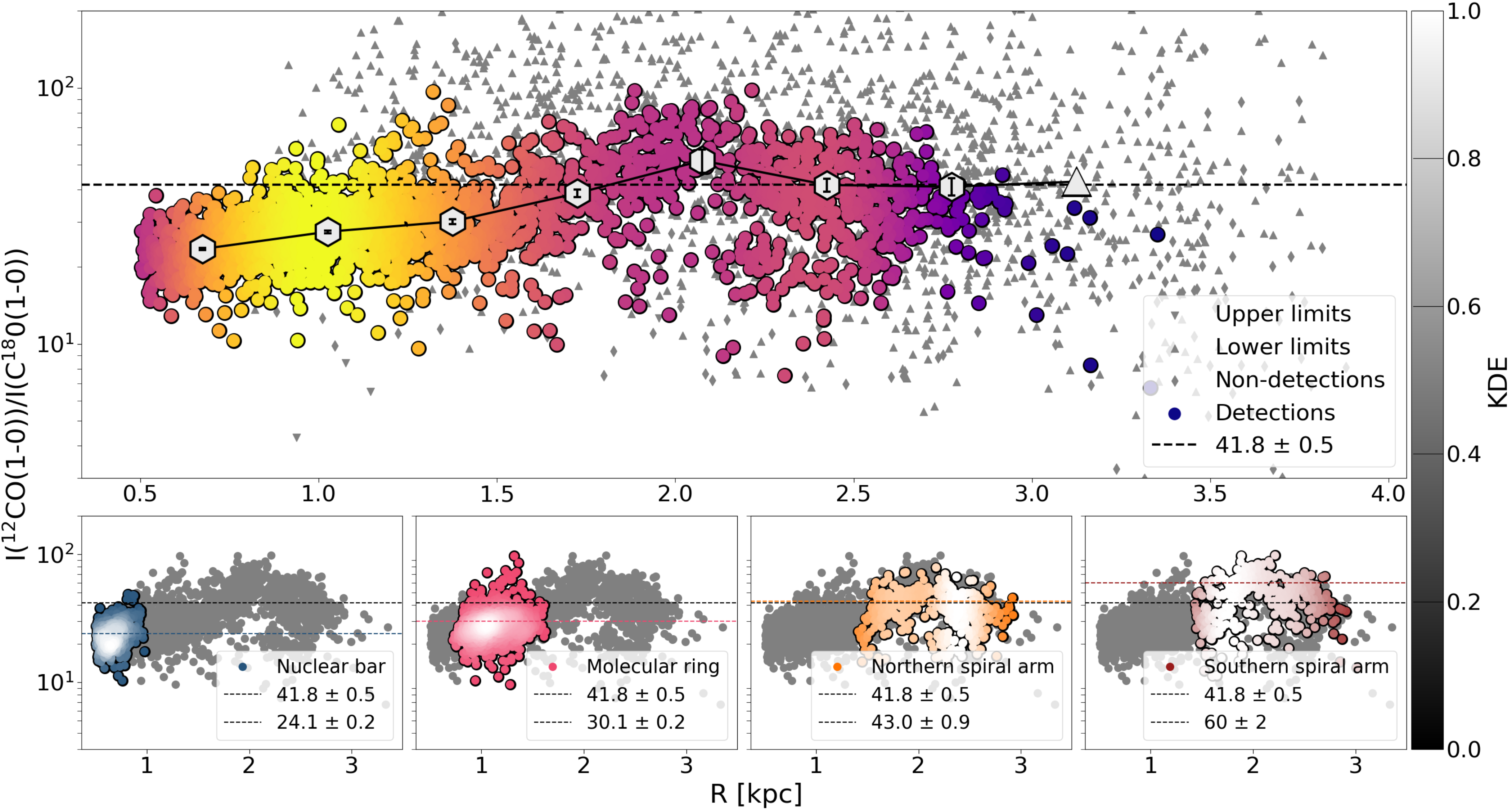}
\caption{The figure displays the $^{12}$CO(1-0) to C$^{18}$O(1-0) line ratio plotted against the galactocentric radius. The description is analogous to that in  Fig.\ref{fig:env_plots_radius}.}
\vspace{5mm}
\label{fig:radius2}
\end{figure*}

\begin{figure*}[h!]
\centering
\includegraphics[width=\textwidth]{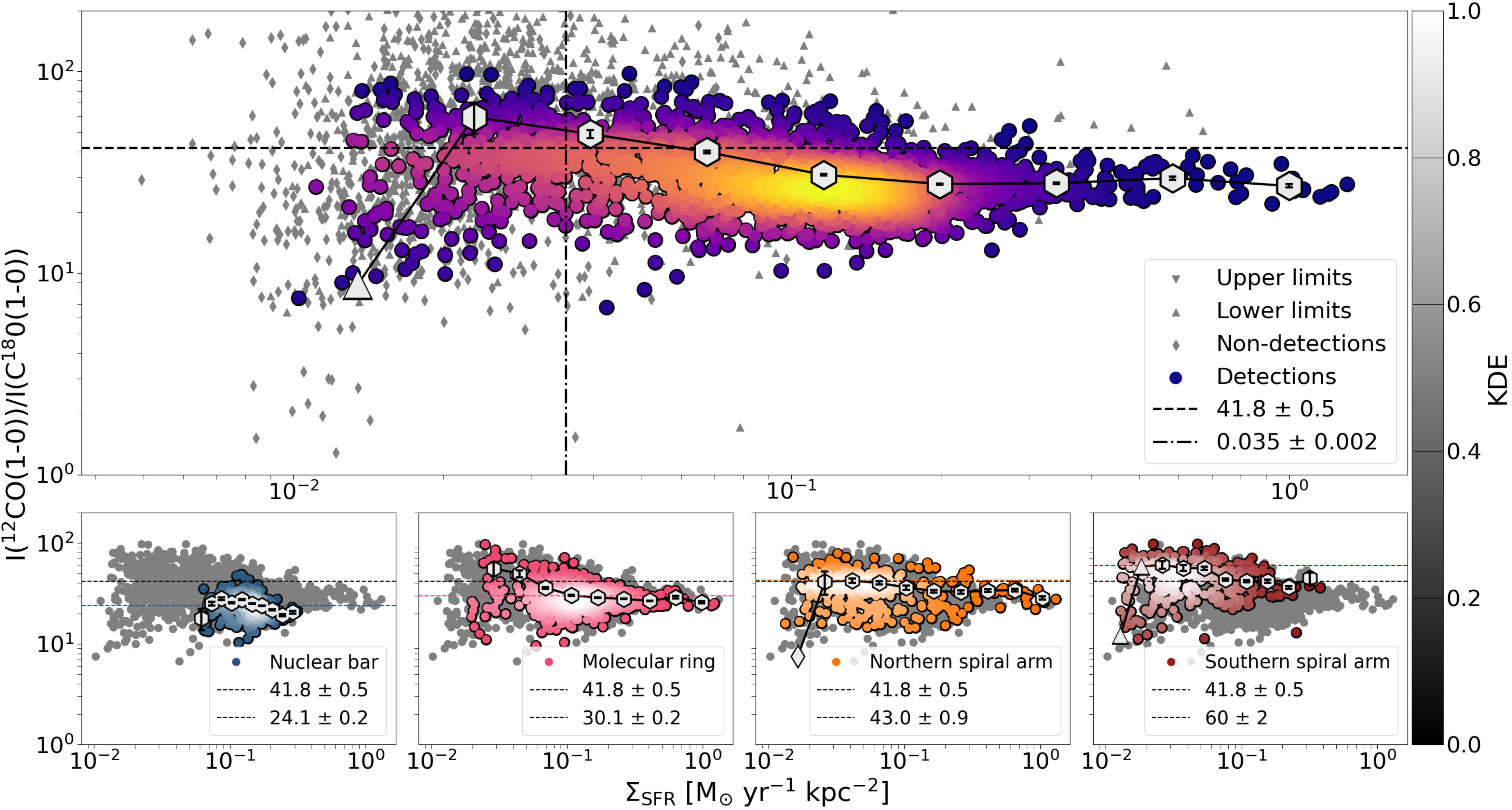}
\caption{The figure displays the $^{12}$CO(1-0) to C$^{18}$O(1-0) line ratio plotted against the $\Sigma_{\text{SFR}}$. The description is analogous to that in  Fig.\ref{fig:env_plots_sfr}.}
\vspace{5mm}
\label{fig:sfr2}
\end{figure*}

\section{Assessing AGN influence on R$^{13}_{18}$}\label{AGN Acivity}
Enhanced UV radiation and cosmic ray fluxes from the AGN may alter the local chemistry, potentially affecting the isotopologue ratios. While $^{13}$CO is more abundant and capable of self-shielding, C$^{18}$O is more easily photodissociated, which could lead to a slight increase in the $^{13}$CO/C$^{18}$O ratio. However, the elevated temperatures in the AGN vicinity likely suppress chemical fractionation processes, which are more efficient in cold gas. Additionally, AGN-driven changes in ion-formation pathways could influence molecular abundances, though the detailed chemistry under such extreme conditions remains uncertain. A more detailed investigation of AGN effects on the SWAN line ratios will be presented in upcoming studies by Usero et al. (in prep.) and Thorp et al. (in prep.).

\onecolumn
\begin{figure}[h!]
\centering
\includegraphics[width=\textwidth]{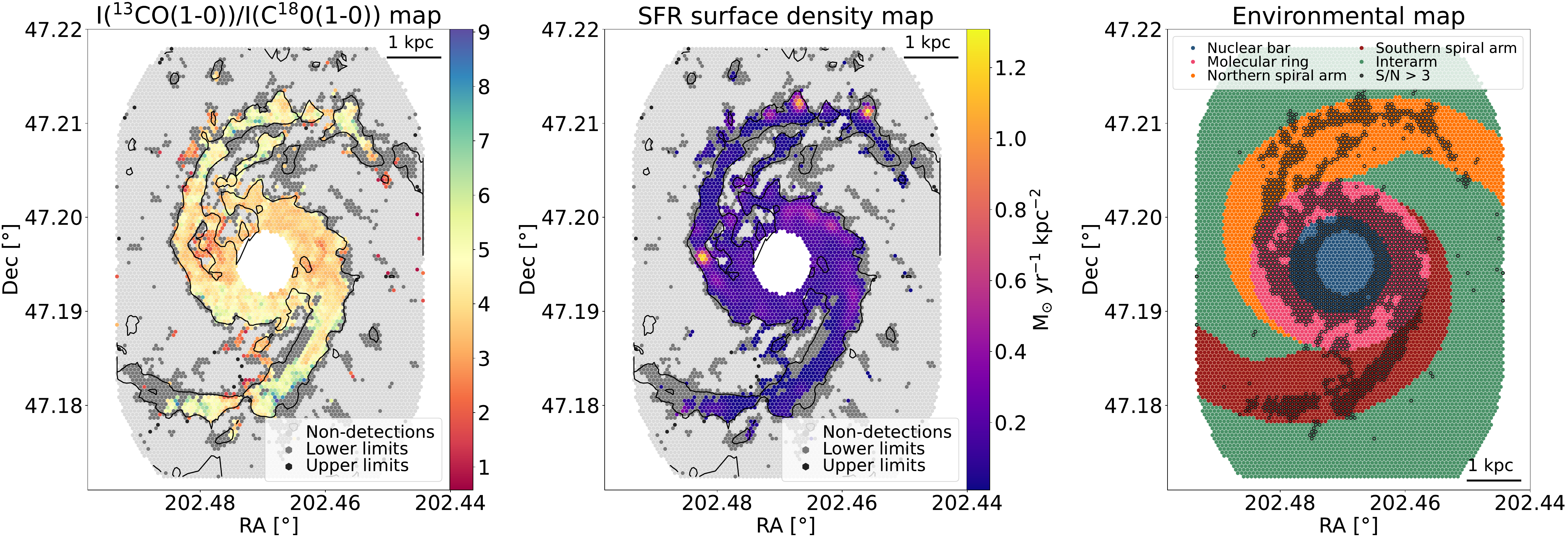}
\caption{The description is analogous to  Fig.\ref{fig:maps}. The white regions in the centre of the galaxy are the regions masked for the AGN activity.}
\label{maps_agn}
\end{figure}

\section{Alternative ratio averaging methods and their uncertainties}\label{Alternative Ratio Averaging Methods and Their Uncertainties}

\begin{table}[h!]
\caption{Comparison of ratio of medians and median of ratios per environment}
\label{table:rom_vs_mor}
\centering
\begin{tabular}{ccccc}
\hline\hline
Region & \makecell{Ratio of \\ medians} & \makecell{Median of \\ ratios with MAD} & 
\makecell{Median of ratios \\ with bootstrapping} & 
\makecell{Relative \\ difference}
 \\
\hline\hline
Whole FoV & 4.34 $\pm$ 0.05 & 3.84 $\pm$ 2 & 3.84 $\pm$ 0.02 & -11\%\\
Nuclear bar & 3.86 $\pm$ 0.03 & 3.88 $\pm$ 0.4 & 3.88 $\pm$ 0.03 & 1\%\\
Molecular ring & 3.89 $\pm$ 0.03 & 3.93 $\pm$ 0.6 & 3.93 $\pm$ 0.02 & 1\%\\
Northern spiral arm & 4.6 $\pm$ 0.1 & 4.47 $\pm$ 2 & 4.47 $\pm$ 0.06 & -3\%\\
Southern spiral arm & 5.4 $\pm$ 0.2 & 4.24 $\pm$ 2 & 4.24 $\pm$ 0.07 & -21\%\\ 
\hline
\end{tabular}
\tablefoot{The relative difference is calculated with respect to the ratio of medians. For instance, in the case of the FoV, the median of ratios is approximately equal to the ratio of medians minus 11\% of its value.}
\end{table}

\begin{multicols}{2}

In Section \ref{Ratio of Medians}, we explained our preference for using the ratio of medians over the median of ratios, primarily due to the impact of C$^{18}$O detectability. The median of ratios method assigns equal weight to each individual ratio which allows low-quality data, specifically pixels with faint C$^{18}$O emission, to disproportionately influence the final result. Nevertheless, for completeness and ease of comparison to other works, we also present the results using the median of ratios method in this section. The median of ratios is computed by first calculating the ratio on a pixel-by-pixel basis, after which the final value is taken as the median of all valid individual ratios:
\begin{equation} 
\overline{R} = \widetilde{\left(\frac{x_i}{y_i} \right)} 
\label{eq4}
\end{equation}
where $x$ and $y$ are integrated intensities of the two tracers (e.g., $^{13}$CO and C$^{18}$O). 
Since we do not censor our data, the propagated uncertainty of the median of ratios exhibits unrealistically high values. To address this, we introduce two alternative methods for estimating the characteristic variation of median of ratios. The first indicator is the median absolute deviation (MAD) of the distribution of pixel-wise ratio uncertainties. For each pixel, the uncertainty in the ratio, $M_{R_i}$, is calculated using standard error propagation on the individual uncertainties $\sigma_{x_i}$ and $\sigma_{y_i}$ of $x_i$ and $y_i$ respectfully:
\begin{equation} 
M_{R_i} = \sqrt{ \left(\frac{\sigma_{x_i}}{y_i} \right)^2 + \left( \frac{x_i}{y_i^2} \, \sigma_{y_i} \right)^2} 
\label{eq5}
\end{equation}
The resulting MADs are listed in the third column of Table \ref{table:rom_vs_mor}, alongside the corresponding medians of ratios for each region. 
The second indicator is obtained through bootstrapping; we generate subsamples of the ratio dataset, allowing repetitions and matching the original sample size. For each subsample, we calculate the median of ratios $\overline{R}$, thereby constructing a distribution of the median of ratios. The MAD of this bootstrapped distribution is then used as a measure of variation and is presented together with the original medians of ratios in the fourth column of Table \ref{table:rom_vs_mor}.
The ratio of medians and the median of ratios values agree well across most environments, with relative differences of only $\approx 1 - 3$\% in the nuclear bar, molecular ring and northern spiral arm. However, the field of view ($\approx-11$\%) and the southern spiral arm ($\approx- 21$\%) show more substantial differences. These discrepancies are still within the uncertainty range of the MAD, but not the tighter bootstrap errors, suggesting there is genuine scatter in the data. This is consistent with what Fig. \ref{fig:delta_maps} shows. The southern spiral arm exhibits some of the extreme values and the full FoV encompasses all environments, including the southern arm, resulting in a broader range of ratios and, consequently, greater discrepancies between the two methods.
\end{multicols}

\end{appendix}

\end{document}